\documentclass{moriond}

\pdfoutput=1
\usepackage{lineno}

\newcommand{\Photo}{}
\begin{document}
\vspace*{4cm}
\title{Light dark matter search with  DarkSide-50}

\author{ D.~Franco$^{1}$ on behalf of the DarkSide-50 Collaboration}

\address{$^{1}$APC, Universit\'e de Paris, CNRS, Astroparticule et Cosmologie, Paris F-75013, France}

\maketitle\abstracts{
We present  the latest results from  the search for light dark matter particle interactions with the DarkSide-50 dual-phase liquid argon time projection chamber. This analysis, based on the  ionization signal only,  improves the existing  limits  for spin-independent WIMP-nucleon interactions  in the $[1.2, 3.6]$~GeV/c$^2$ mass range.  The sensitivity is extended down to~40~MeV/c$^2$ by assuming the Migdal effect, responsible  for an additional ionization signal from the recoiling atom. Finally, we set new constraints to interactions of dark matter particles with  electrons in the final state, namely WIMPs, galactic axions, dark photons, and sterile neutrinos. 
}

\section{Introduction}
DarkSide-50 was a dual-phase liquid argon (LAr) time projection chamber (TPC), operational between 2013 and 2019 in the Hall C of the Gran Sasso National Laboratory (LNGS) in Italy, primarily designed to  search for  spin-independent interactions between nucleons and high-mass WIMPs  ($>$10 GeV/c$^2$). The first data-taking campaign ran from November 2013 to April 2015 with an atmospheric argon (AAr) target, contaminated  with cosmogenic $^{39}$Ar with a specific activity of $\sim$1~Bq/kg.  The target was  then replaced with low-radioactivity argon extracted from deep underground (UAr), where activation of cosmogenic isotopes is highly suppressed. 

The   active mass,  46.4$\pm$0.7 kg, is enclosed in  a PTFE cylinder on the side  and by  two arrays of nineteen 3'' photomultiplier tubes  (PMTs)  on the top and bottom. The PMTs observe  scintillation and electroluminescence light pulses, induced by particle interactions in LAr. The scintillation pulse originates from atomic excitation and  recombination of ion-electron pairs produced by ionization. The electroluminescence one is induced by ionization electrons, which survive  the recombination. These are drifted upwards,  across the liquid bulk, thanks to a uniform 200 V/cm electric field,  shaped by copper rings surrounding the PTFE sidewalls. Ionization electrons are  then extracted and drifted in gaseous argon at the top of the TPC, where they produce, in average, $\sim$23 photoelectrons per electron extracted in gas.


All the inner surfaces of the TPC are coated with tetraphenyl butadiene (TPB), a wavelength shifter that absorbs 128 nm photons from argon de-excitation and re-emits photons, whose wavelengths are peaked at 420 nm.  The TPC, hosted inside a 120 l cryostat,  is shielded against  neutrons and cosmic muons by  surrounding liquid scintillator and water Cherenkov detectors. 

The  discovery potential of dual-phase LAr TPCs in searching for high-mass ($>$10~GeV/c$^2$) WIMPs  relies on their extraordinary background discrimination. This was proved by DarkSide-50 with the AAr campaign, by rejecting  1.5$\times$10$^7$  electronic recoils from $^{39}$Ar decays with the scintillation pulse shape discrimination~\cite{DarkSide:2014llq}. The potential was later confirmed with  the UAr dataset, where no  event was observed in the acceptance region for WIMPs  over  532 days of live search~\cite{DarkSide:2018kuk}. 

In 2018,  DarkSide-50  extended its physics case to lighter dark matter particles~\cite{DarkSide:2018bpj,DarkSide:2018ppu} by lowering the energy threshold from $\sim$20~keV$_{nr}$ (nuclear recoil equivalent)  to few hundreds of eV$_{nr}$.  This was achieved by using  the ionization-based event energy estimator instead of  the scintillation one. Single ionization electrons are detected with 27\% resolution  and with nearly 100\%  extraction  efficiency  in gas, to be compared with 16\%, the efficiency in detecting a scintillation photon~\cite{DarkSide:2017wdu}.  The drawback  is the reduced background discrimination power due to the absence of scintillation pulse shape and volume fiducialization along the electric field, as the drift time is no longer measurable. Nevertheless,  this analysis technique, applied for the first time to a liquid argon experiment, led in 2018 to the  improvement of existing limits on WIMP-nucleon and WIMP-electron interactions in the few GeV/c$^2$ and sub-GeV/c$^2$ mass ranges, respectively~\cite{DarkSide:2018bpj,DarkSide:2018ppu}. 

The analysis presented here is based on 653.1 live-days from the UAr campaign from December 12, 2015, to February 24, 2018, nearly twice the exposure used in the analysis published in 2018~\cite{DarkSide:2018bpj}. The new analysis profits  from  an improved data selection and a more accurate background model. Moreover, it benefits from the extensive effort devoted in 2021 to accurately determine the LAr ionization response  to nuclear (NR) and electronic (ER) recoils, down to the sub-keV range~\cite{DarkSide:2021bnz}.   The  ER ionization response was measured  down to $\sim$180 eV$_{er}$, exploiting $^{37}$Ar and $^{39}$Ar decays naturally present in the early LAr dataset, and extrapolated to a 3 ionization electrons with the Thomas-Imel box model~\cite{Thomas:1987zz}. The ionization response to NRs was measured down to $\sim$500~eV$_{nr}$, the lowest ever achieved in liquid argon, using \textit{in situ} neutron calibration sources and external datasets from neutron beam experiments, like ARIS~\cite{Agnes:2018mvl} and SCENE~\cite{SCENE:2014iyj}. Other elements of the re-analysis are an improved detector response model and a refined treatment of systematics into the statistical analysis.

\section{Data selection and background model}

Events are required  for this analysis to be single-scatter, hence selected  with a single reconstructed ionization pulse (S2).  A set of quality cuts are applied in order to reject unresolved pile-ups of scintillation-ionization or multiple-ionization pulses. Quality cuts rely on   pulse shape estimators like the peak and the FWHM of the S2 time profile, and  the pulse start time with respect to the  offset of the trigger, which requires at least two PMTs above a threshold of 0.6 PE. Single scatter events detected by the outermost ring of PMTs, mostly  due to external radioactive contamination,  are discarded, reducing the signal acceptance to 41.9\%. 

\begin{figure}
\centerline{\includegraphics[width=0.6\linewidth]{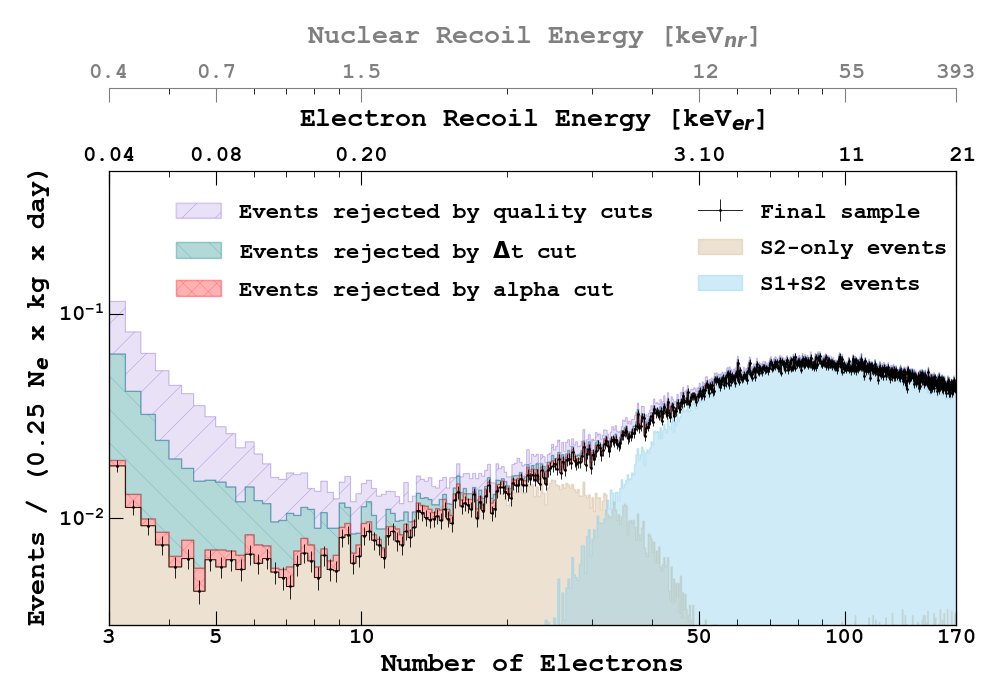}}
\caption[]{Ionization electron spectra at different steps of the data selection, after rejection of events outside the fiducial volume and with multiple interactions~\cite{DarkSide:2022knj}.}
\label{fig:bg}
\end{figure}

The  resulting data sample  is contaminated by two classes of background events that can be rejected on an event-by-event basis. The first  is associated to spurious ionization electrons. These are trapped along their drift by trace impurities or at the liquid surface and released with delays of up to hundreds of milliseconds. This background component is partially suppressed by applying a  veto of 20 ms after each DAQ trigger.  The  second class  is associated to events with a large scintillation component and an anomalously low S2 pulse. Their origin is associated to $\alpha$ particles generated  at the TPC walls, where ionization electrons are suppressed being absorbed by the walls  themselves. Some of the scintillation photons may extract  electrons from the cathode by photoelectric effect inducing low S2 pulses. These events are rejected on the basis of the ionization-to-scintillation ratio, a cut tuned on ER and NR calibration datasets.  The final dataset   corresponds to an exposure of (12,306 $\pm$184) kg d. It contains about 300,000 events in [4,170] $N_e$ (number of electrons), the region of interest (RoI) for this analysis, equivalent to [0.06, 21] keV$_{er}$ ([0.64, 288] keV$_{nr}$) in the ER (NR) energy scale~\cite{DarkSide-50:2022qzh}. The impact of each step of the data selection is shown in Figure~\ref{fig:bg}. 

The selected data sample is dominated by background events from radioactive  contamination in LAr (internal)  and  in the detector components surrounding the active target (external). Internal background is mainly due to  first forbidden unique beta decays of $^{39}$Ar and $^{85}$Kr, whose activities in the RoI  and in the fiducial volume is estimated to be (6.5$\pm$0.9)$\times$10$^{-4}$~Bq and (1.7$\pm$0.1)$\times$10$^{-3}$~Bq, respectively. The associated theoretical energy spectra  used in this analysis take into account recent calculations of atomic exchange and screening effects~\cite{PhysRevA.90.012501}. The external background is mostly due to   $\gamma$s and X-rays from detector components, whose specific activities were determined via a comprehensive material screening campaign. As the spectral shapes of backgrounds originating from nearby positions resemble each other  in the RoI, they were combined into two main components,  here labelled as   PMTs and  cryostat. Backgrounds  from radiogenic and cosmogenic neutrons, as well as coherent elastic neutrino-nucleus scattering from solar and atmospheric neutrinos, are negligible and hence not included in this analysis.

\begin{figure}
\centerline{\includegraphics[width=0.65\linewidth]{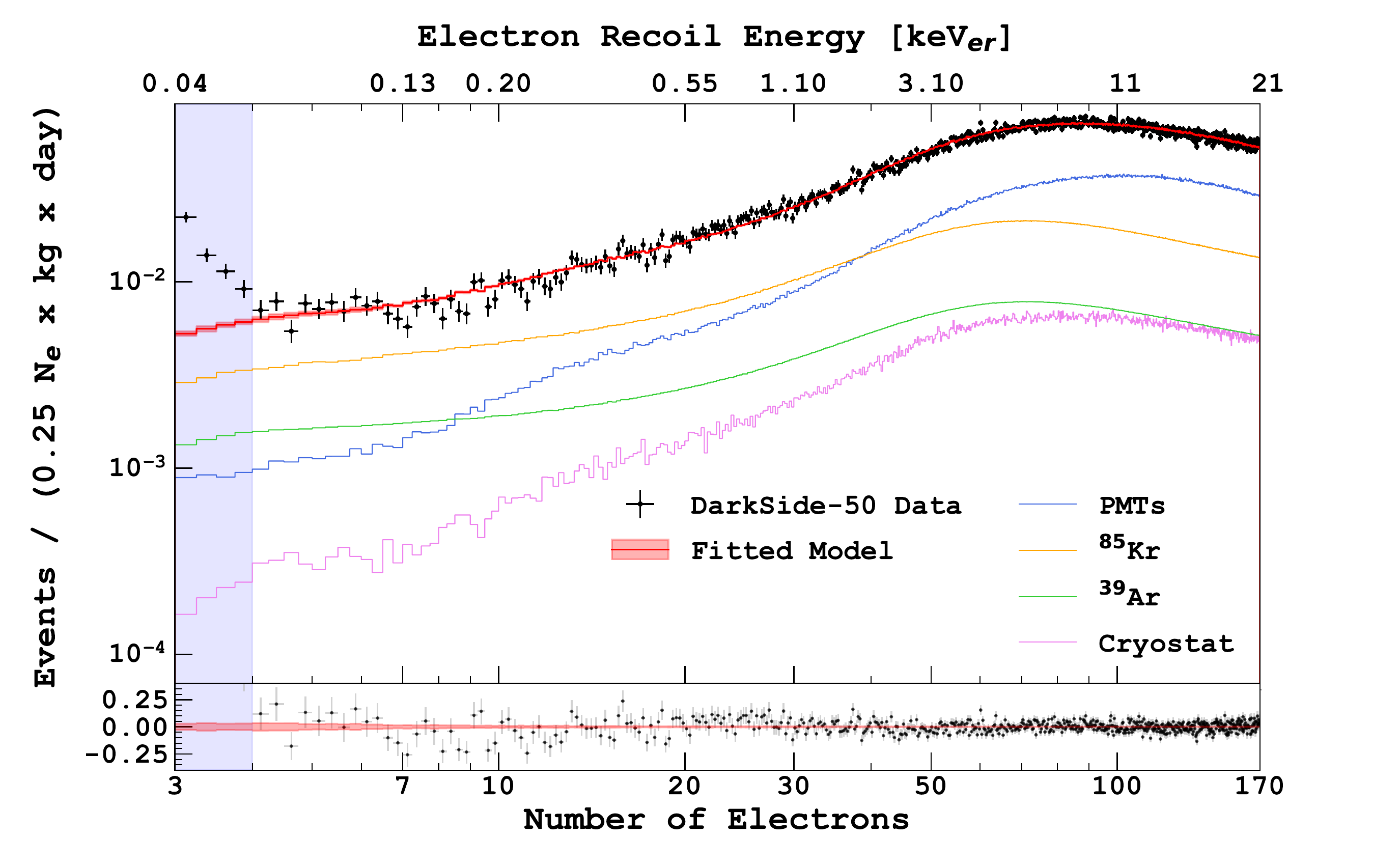}}
\caption[]{Fit of the background-only model   to the DarkSide-50 data, after selection cuts in the [4, 170]~Ne range.  The associated  uncertainty (shaded area), residuals, and the  individual contributions  from the internal ($^{39}$Ar and $^{85}$Kr) and external radioactive backgrounds (cryostat and PMTs) are also shown~\cite{DarkSide:2022knj}.}
\label{fig:fitbg}
\end{figure}

The background model is built by simulating each isotope,  uniformly distributed in the associated component materials. Decaying particles are tracked over the DarkSide-50 geometry with  G4DS, the DarkSide Geant4-based Monte Carlo~\cite{DarkSide:2017wdu}.   The simulation of the detector response  accounts for, among other effects, electron captures by impurities along the drift,   the dependence of the S2 response on the event radial position, and the energy dependent inefficiencies from the quality cuts.  

Data are analyzed with a binned Profile Likelihood Ratio approach, which accounts for systematics by means of 11 nuisance parameters,  classified as ``amplitude'', \textit{i.e.}, acting on the   normalization  of the background components, and as ``shape'', accounting for  spectral distortions from the ionization response and from  uncertainties on  $^{39}$Ar and $^{85}$Kr $\beta$-decay spectral shapes~\cite{DarkSide-50:2022qzh}. The data fit with the background-only model is shown in Fig.~\ref{fig:fitbg}. The excess below the analysis threshold of 4 e$^-$ is associated to the presence of  spurious electrons.

\begin{figure}
\includegraphics[width=0.48\linewidth]{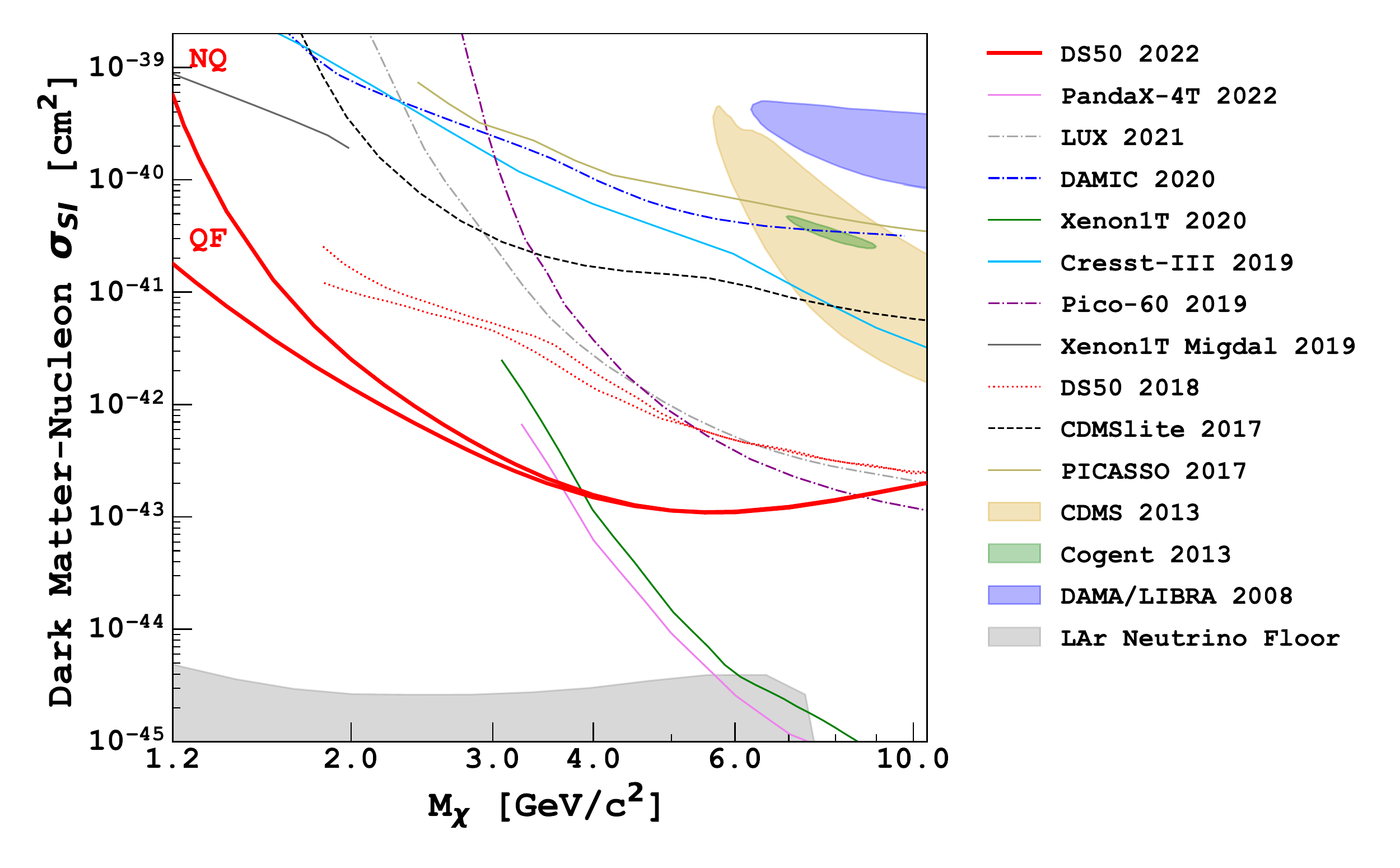}
\includegraphics[width=0.48\linewidth]{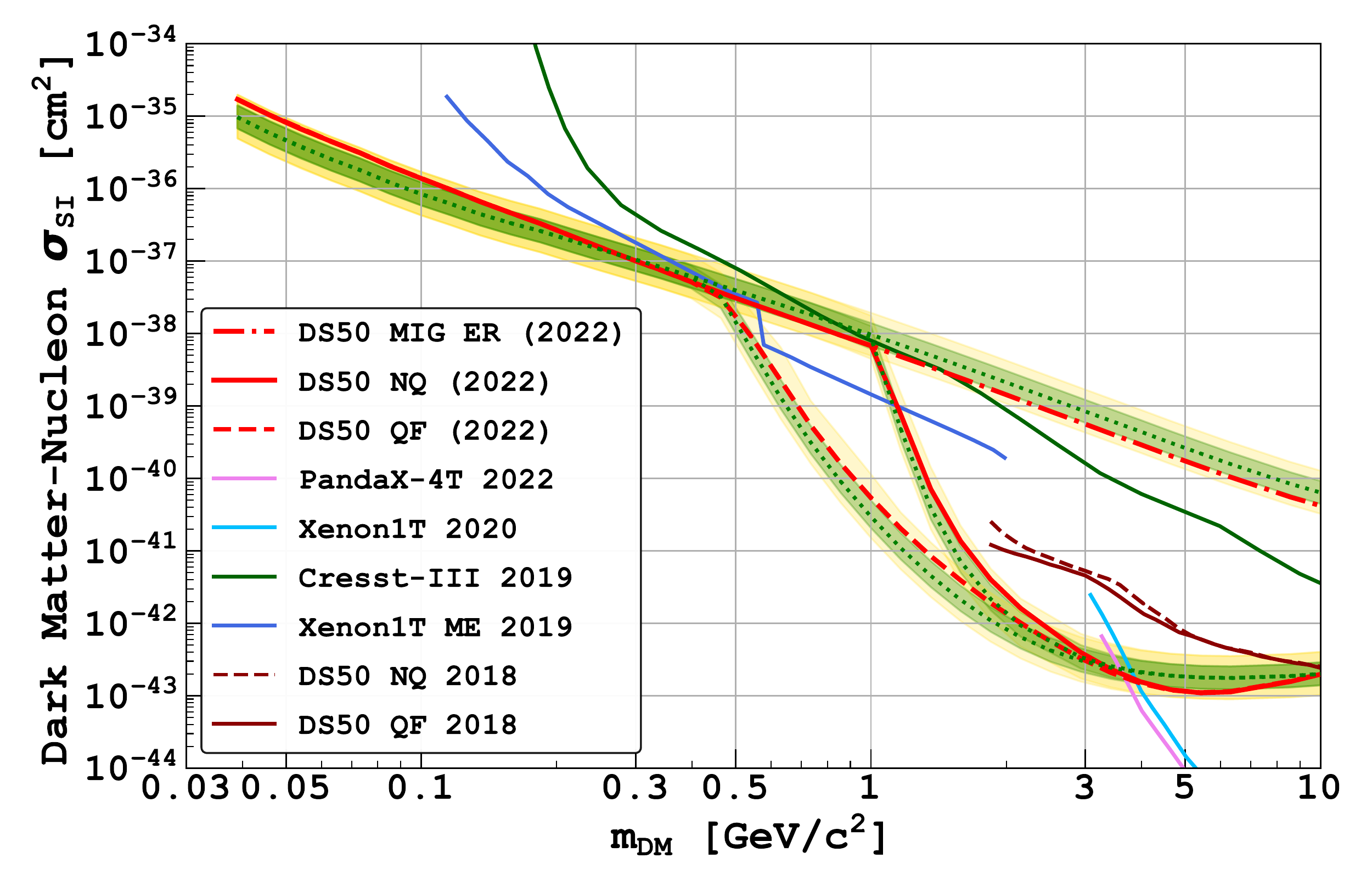}
\caption[]{Left. DarkSide-50 limits with (QF) and without (NQ) quenching fluctuations, compared to existing 90\% C.L. exclusion limits and claimed discovery,  and to the neutrino floor for LAr experiments~\cite{DarkSide:2022knj}. Right. DarkSide-50 limits with (QF) and without (NQ) quenching fluctuations, extended down to 40~MeV/c$^2$ WIMP mass by exploiting the Migdal effect~\cite{DarkSide:2022dhx}.}
\label{fig:wimp}
\end{figure}

\begin{figure}
\centerline{\includegraphics[width=0.7\linewidth]{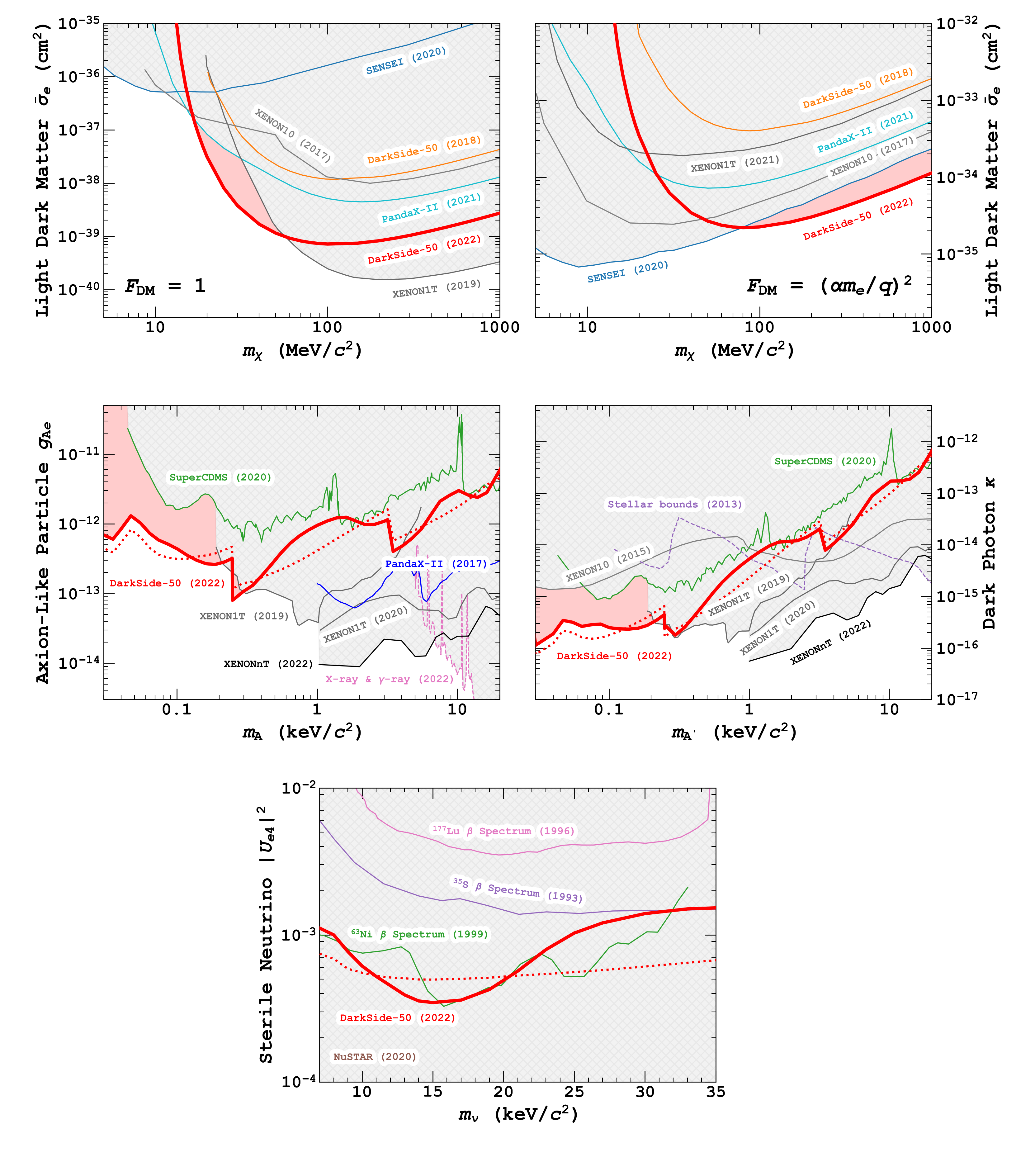}}
\caption[]{Exclusion limits at 90\% C.L. on DM particle interactions with electron final states~\cite{DarkSide-50:2022qzh}.}
\label{fig:lepto}
\end{figure}

\section{Results and conclusions}

The NR signals from WIMP interactions via elastic scattering  are modeled with the same Monte Carlo approach used for the background components.  The systematic sources affecting the signal model arise from the NR ionization response uncertainty  and from  the fiducial volume one (1.5\%), accounted for  in the likelihood by \textit{shape} and  \textit{amplitude} nuisance parameters, respectively.  The only model uncertainty that cannot be parameterized in the likelihood, due to the lack of an accurate description of the process, is related to the fluctuation induced by the ionization quenching effect.  This plays a major role as, in addition to fluctuations resulting from the  excitons and ionization electrons partition and from the ion-electron recombination, it increases the probability of observing events above the analysis threshold. We therefore assumed two cases. The first is the suppression of quenching fluctuations (NQ), which, although not physical, represents  the most conservative modelling with respect to the WIMP search. In the second,  fluctuations are assumed to be governed by binomial statistics (QF), between detectable (ionization electrons and excitons) and undetectable quanta (e.g. phonons).

Exclusion limits above 1.2~GeV/c$^2$ WIMP mass are shown  in Fig.~\ref{fig:wimp} (left). Assuming  NQ fluctuations, the most conservative model, DarkSide-50 establishes the current  best 90\% C.L. limits for WIMPs with masses in the range [1.2, 3.6]~GeV/c$^2$ and improves  by a factor of $\sim$10 at 3~GeV/c$^2$ the results obtained in 2018~\cite{DarkSide-50:2022qzh}.  The limit was then extended down to 40~MeV/c$^2$ by taking into account the Migdal effect~\cite{DarkSide:2022knj}, responsible of  extra ionization or excitation components emitted because of the instantaneous momentum change, with respect to the atomic electrons, of the nucleus elastically scattered at rest. As shown in   Fig.~\ref{fig:wimp} (right), the limit is entirely driven by the Migdal effect for WIMP masses  below 0.5~GeV/c$^2$. This results was confirmed by  an alternative analysis based on Bayesian Networks \cite{DarkSide-50:2023fcw}.

As for \textit{leptophilic} dark matter~\cite{DarkSide:2022dhx}, \textit{i.e.}, interacting with an electron in the final state, we have established the best direct-detection 90\% C.L. limits on dark matter-electron scattering in the [16, 56]~MeV/c$^2$ mass range  for a heavy mediator and above 80~MeV/c$^2$ for a light mediator. We have also placed the first constraints on galactic axion-like particles and dark photons. Moreover, DarkSide-50 is the first direct dark matter direct-detection experiment to set limits on the sterile neutrino mixing angle. All the leptophilic limits are shown in Fig.~\ref{fig:lepto}.

These results may be improved in the future by better defining the LAr ionization response and the stochastic model underlying the NR quenching. In addition, they demonstrate the strong potential of dual-phase LAr TPC technology, with promising prospects from DarkSide-20k, the next generation 50-ton LAr TPC under construction at LNGS~\cite{DarkSide-20k:2017zyg}. 

\section*{Acknowledgments}

This work was partially supported  by the UnivEarthS LabEx program (Grants No. ANR-10-LABX-0023 and No. ANR-18-IDEX-0001). 

\section*{Bibliography}
\bibliography{biblio}

\begin{thebibliography}{10}

\bibitem{DarkSide:2014llq}
P.~Agnes et~al.
\newblock {First Results from the DarkSide-50 Dark Matter Experiment at
  Laboratori Nazionali del Gran Sasso}.
\newblock {\em Phys. Lett. B}, 743:456--466, 2015.

\bibitem{DarkSide:2018kuk}
P.~Agnes et~al.
\newblock {DarkSide-50 532-day Dark Matter Search with Low-Radioactivity
  Argon}.
\newblock {\em Phys. Rev. D}, 98(10):102006, 2018.

\bibitem{DarkSide:2018bpj}
P.~Agnes et~al.
\newblock {Low-Mass Dark Matter Search with the DarkSide-50 Experiment}.
\newblock {\em Phys. Rev. Lett.}, 121(8):081307, 2018.

\bibitem{DarkSide:2018ppu}
P.~Agnes et~al.
\newblock {Constraints on Sub-GeV Dark-Matter\textendash{}Electron Scattering
  from the DarkSide-50 Experiment}.
\newblock {\em Phys. Rev. Lett.}, 121(11):111303, 2018.

\bibitem{DarkSide:2017wdu}
P.~Agnes et~al.
\newblock {Simulation of argon response and light detection in the DarkSide-50
  dual phase TPC}.
\newblock {\em JINST}, 12(10):P10015, 2017.

\bibitem{DarkSide:2021bnz}
P.~Agnes et~al.
\newblock {Calibration of the liquid argon ionization response to low energy
  electronic and nuclear recoils with DarkSide-50}.
\newblock {\em Phys. Rev. D}, 104(8):082005, 2021.

\bibitem{Thomas:1987zz}
J.~Thomas and D.~A. Imel.
\newblock {Recombination of electron-ion pairs in liquid argon and liquid
  xenon}.
\newblock {\em Phys. Rev. A}, 36:614--616, 1987.

\bibitem{Agnes:2018mvl}
P.~Agnes et~al.
\newblock {Measurement of the liquid argon energy response to nuclear and
  electronic recoils}.
\newblock {\em Phys. Rev. D}, 97(11):112005, 2018.

\bibitem{SCENE:2014iyj}
H.~Cao et~al.
\newblock {Measurement of Scintillation and Ionization Yield and Scintillation
  Pulse Shape from Nuclear Recoils in Liquid Argon}.
\newblock {\em Phys. Rev. D}, 91:092007, 2015.

\bibitem{DarkSide-50:2022qzh}
P.~Agnes et~al.
\newblock {Search for low-mass dark matter WIMPs with 12~ton-day exposure of
  DarkSide-50}.
\newblock {\em Phys. Rev. D}, 107(6):063001, 2023.

\bibitem{DarkSide:2022knj}
P.~Agnes et~al.
\newblock {Search for Dark Matter Particle Interactions with Electron Final
  States with DarkSide-50}.
\newblock {\em Phys. Rev. Lett.}, 130(10):101002, 2023.

\bibitem{PhysRevA.90.012501}
X.~Mougeot and C.~Bisch.
\newblock Consistent calculation of the screening and exchange effects in
  allowed ${\ensuremath{\beta}}^{\ensuremath{-}}$ transitions.
\newblock {\em Phys. Rev. A}, 90:012501, Jul 2014.

\bibitem{DarkSide:2022dhx}
P.~Agnes et~al.
\newblock {Search for Dark-Matter\textendash{}Nucleon Interactions via Migdal
  Effect with DarkSide-50}.
\newblock {\em Phys. Rev. Lett.}, 130(10):101001, 2023.

\bibitem{DarkSide-50:2023fcw}
P.~Agnes et~al.
\newblock {Search for low mass dark matter in DarkSide-50: the bayesian network
  approach}.
\newblock 2 2023.

\bibitem{DarkSide-20k:2017zyg}
C.~E. Aalseth et~al.
\newblock {DarkSide-20k: A 20 tonne two-phase LAr TPC for direct dark matter
  detection at LNGS}.
\newblock {\em Eur. Phys. J. Plus}, 133:131, 2018.

\end{thebibliography}

\end{document}